\documentclass[9pt,journal,letterpaper]{IEEEtran} 

\usepackage{setspace}
\usepackage{lipsum}  

\usepackage{makecell}
\setcellgapes{2pt}

\usepackage{textcase}
\usepackage[tablename=TABLE]{caption}
\DeclareCaptionTextFormat{up}{\MakeTextUppercase{#1}}
\usepackage{graphicx}
\usepackage{multirow}

\usepackage{float}
\usepackage{amsthm}
\usepackage{commath}
\usepackage{array}
\usepackage{booktabs}

\DeclareMathOperator{\sir}{SIR}

\usepackage{bbm}
\usepackage{amssymb}
\usepackage{cite}

\usepackage{subcaption}
\usepackage[labelformat=simple]{subcaption}

\graphicspath{{images/}}

\newtheorem{thm}{Theorem}

\newtheorem{definition}{Definition}

\allowdisplaybreaks

\newcommand{\quotes}[1]{``#1''}

\begin{document}

	\title{Optimal Coverage and Rate in Downlink Cellular Networks: A SIR Meta-Distribution Based Approach}
	
	
	\author{A. M. Hayajneh, S. A. R. Zaidi, Des. C. McLernon, M. Z. Win, A. Imran and M. Ghogho
		\thanks{ This work is accepted and to appear at 2018 IEEE Global Communications Conference: Mobile and Wireless Networks at: Abu Dhabi, UAE.}
		
 \thanks{This work is partially funded by the Hashemite University (HU), Zarqa, Jordan.} \thanks{This work was also partly supported by the UK British Council (Newton Fund) through the Project \quotes{Wireless Sensor Networks for Real Time Monitoring of Water Quality} under Grant IL3264631003.}
 \thanks{A. M. Hayajneh, S. A. R. Zaidi and Des. C. McLernon and M. Ghogho are with the School of Electronic and Electrical Engineering, University of Leeds, Leeds LS2 9JT, United Kingdom, e-mail: \{elamh,s.a.zaidi,d.c.mclernon\}@leeds.ac.uk. A. M. Hayajneh is also affiliated with the Hashemite University, Zarqa, Jordan. M. Ghogho is also with the International University of Rabat, TICLab, Technopark Rabat-Salé, 11100, Morocco, e-mail: m.ghogho@ieee.org. M.Z. Win is with the Laboratory for Information and Decision Systems, Massachusetts Institute of Technology, Cambridge, MA 02139 USA, e-mail: moewin@mit.edu. A. Imran is with the School of Electrical and Computer Engineering, The University of Oklahoma, Tulsa, OK 74104, USA, e-mail: ali.imran@ou.edu.}}
 
 \maketitle
	
	
	\begin{abstract}
	In this paper, we present a detailed analysis of the coverage and spectral efficiency of a downlink cellular network. Rather than relying on the first order statistics of received signal-to-interference-ratio (SIR) such as coverage probability, we focus on characterizing its meta-distribution. Our analysis is based on the alpha-beta-gamma (ABG) path-loss model which provides us with the flexibility to analyze urban macro (UMa) and urban micro (UMi) deployments. With the help of an analytical framework, we demonstrate that selection of underlying degrees-of-freedom such as BS height for optimization of first order statistics such as coverage probability is not optimal in the network-wide sense. Consequently, the SIR meta-distribution must be employed to select appropriate operational points which will ensure consistent user experiences across the network. Our design framework reveals that the traditional results which advocate lowering of BS heights or even optimal selection of BS height do not yield consistent service experience across users. By employing the developed framework we also demonstrate how available spectral resources in terms of time slots/channel partitions can be optimized by considering the meta-distribution of the SIR.
	\end{abstract}
	

\section{Introduction}

\subsection{Motivation and Related Work}
Network densification is considered as a key design tool to satisfy the ever-increasing demand for any-time, anywhere wireless connectivity. The fundamental idea behind densification is to bring the network closer to the user, i.e. reduce the cell size while increasing the density of deployment. Fundamentally, this enables more aggressive spectrum reuse across spatial dimensions, resulting in enhanced network throughput. Moreover, reduction in cell-size results in an improved coverage for the intended users. Nevertheless, the aggressive spatial reuse, when coupled with reduced cell size, poses significant challenges in terms of interference management. This has resulted in a fundamental question which has intrigued network designers and researchers for past few years: “How are coverage and throughput related to the deployment density?”

To answer this question, the performance of large-scale cellular networks must be quantified in terms of underlying design parameters such as base station (BS) density, path-loss exponents, transmit power employed by BSs and available channel resources etc. Unfortunately, traditional analysis based on hexagonal tessellation does not yield any significant insight due to lack of analytical tractability. In \cite{andrews2011tractable}, the authors introduced a tractable approach for the analysis of coverage and rate in large-scale wireless networks using stochastic geometry. In the recent past \cite{wang2018sir,elsawy2017modeling, kalamkar2017spatial,multi_slope_4_comprehensive,haenggi2016meta,ding2016please, haenggi_70_pound, wang2017meta, hayajneh2017performance,hayajneh2018performance, deng2017fine,win2006error}, stochastic geometry has been extensively employed to investigate the design space of large-scale cellular deployments under different 5G architectures and access methodologies. The interested reader is referred to \cite{elsawy2017modeling} for a detailed survey.

The results in \cite{andrews2011tractable} demonstrated that the coverage probability in a signal to interference ratio (SIR) limited scenario is independent of the BSs density. Consequently, the network area spectral efficiency increases with an increase in the BS density. The analysis was revisited in \cite{multi_slope_4_comprehensive} who investigated the impact of line of sight (LoS) and non-line of sight (NLoS) propagation on the coverage and area spectral efficiency. The authors demonstrated that in contrast to \cite{andrews2011tractable}, there exists an optimal BS density beyond which the area spectral efficiency is reduced with further densification. Both of these analyses ignored the impact of BS height on the network performance. In \cite{ding2016please}, the authors extended the analysis of \cite{multi_slope_4_comprehensive} to capture the impact of non-zero height difference between user equipment (UE) and BS. The authors presented area spectral efficiency (ASE) crash, i.e. the phenomenon of significant deterioration in ASE with network densification with realistic elevation consideration. This framework is further extended in \cite{atzeni2018downlink} and \cite{multi_slop} under different fading considerations. One of the key observations which follows from these studies is that the BS heights should be lowered as it reduces the path-loss between UE and BS. On a closer inspection it is obvious that the path-loss model used in \cite{atzeni2018downlink} and \cite{multi_slop} does not adequately capture the fact that LoS probability increases with an increase in the BS height along with the path-loss and NLoS probability increases with a decrease in BS height while the path-loss also decreases with a lowering of the BS height. This is indeed adequately captured in \cite{access2010further} and is employed in \cite{al2014optimal} to investigate optimal height for a low altitude platform empowered with a cellular BS. Consequently, it is obvious that there exists an optimal height for the BS deployment which will maximize the area spectral efficiency and lowering the BS height is not always optimal.

All these investigations are based on first-order analysis, i.e. on the coverage probability. In \cite{haenggi2016meta} the authors showed that although the calculation of SIR distribution for the cellular network (which provides a basis for coverage and spectral efficiency calculations) is straight-forward, it only provides limited information about the coverage of individual links. In other words, it is difficult to establish what percentage of links will be able to experience a certain coverage for a desired target SIR threshold from the coverage probability alone. The authors in \cite{haenggi2016meta} presented a framework for the evaluation of what is known as the meta-distribution of SIR which is the distribution of coverage conditional on a point process (see section III for further details). The meta-distribution of the SIR is a better metric as averaging can be often misleading. Notice that the meta-distribution in \cite{ haenggi2016meta} is for the power-law path-loss model which does not discriminate between NLoS and LoS propagation. Combining insights from \cite{atzeni2018downlink,multi_slop} and \cite{haenggi2016meta} one may ask a really important design question, i.e., if $h^*$ is the BS height which maximizes the network performance on average, i.e. in terms of coverage probability for instance, does it also minimize the variance in coverage? In other words, is the $h^*$ which maximizes first order performance metric optimal in terms of the SIR meta-distribution. To this end, this paper presents a comprehensive framework for the investigation of the design space of large-scale cellular networks in terms of the meta-distribution of SIR considering the realistic propagation model.

\subsection{Contributions}
The key contributions of this paper are as follows:
\begin{enumerate}
\item Considering a very general ABG propagation model, we first quantify the coverage probability and rate coverage probability for downlink communications in a large-scale cellular network.
\item	We then present an analytical framework to quantify higher-order moments of the coverage and rate coverage probability which quantify the respective meta-distribution. 
\item	The meta-distribution of SIR is often recovered using higher-order moments in conjunction with the Gil-Pelaez theorem \cite{haenggi2016meta}. However, this requires complex integration for which numerical integration takes a long time to converge. We present a solution based on Mnatsakanov's theorem which simplifies and speed up the evaluation of the meta-distribution.
\item	We investigate the design space of the considered network and present several important insights. 
\item	Lastly, we consider the resource allocation problem in terms of bandwidth partitioning or time-slot sharing and show how such a problem can be tackled using the meta-distribution.
\end{enumerate}

\subsection{Organization}
The rest of the paper is organized as follows: Section \ref{sec:system_model} introduces the system model and deployment geometry of the network. Section \ref{sec:analysis} gives the performance analysis and mathematical modelling. Section \ref{discussion} presents numerical results. Finally, Section \ref{sec:conclusion} provides some future work and conclusions. 
\subsection{Notations}
The probability density function (PDF) for a random variable $X$ is represented as $f_X(x)$ with the cumulative density function written as $F_X(x)$. The exclusion symbol $\setminus$ represents the exclusion of a subset from a superset. The expectation of a function $g(X)$ of a random variable $X$ is represented as $\mathbb{E}_X[g(X)]$. The bold-face lower case letters (e.g., {$\bf x$}) are employed to denote a vector in $\mathbb{R}^2$ and $\norm{\bf x}$ is its Euclidean norm.
\section{System model} \label{sec:system_model}
\begin{table}[]
	\makegapedcells
	\caption{Path loss model parameters}
	\label{table:pathloss_table}
	\centering
	\begin{tabular}{clrlll}
		\toprule			\toprule
		\multicolumn{1}{l}{} Type & Model & $\lambda$ range   & $\alpha$ & $\beta$ & $\gamma$ \\ \midrule\midrule
		\multirow{2}{*}{UMi} & LoS         & $10^{-3}- 10^{-5}$&  2.0    & 31.4     &  2.1        \\ 
		& NLoS        & $10^{-3}- 10^{-5}$&  3.5    & 24.4     &  1.9        \\ \midrule
		\multirow{2}{*}{UMa} & LoS         &$10^{-5}- 10^{-7}$ &  2.8    & 11.4     &  2.3        \\ 
		& NLoS        &$10^{-5}- 10^{-7}$ &  3.3    &  17.6    &  2.0        \\\bottomrule\bottomrule
	\end{tabular}   
\end{table} 
{\it  Spatial and Network Models:} We consider a large-scale cellular network where the locations of the BSs are modelled by homogeneous Poisson point process HPPP such that \cite{stoyan}:
\begin{equation}
 \Phi= \{{\bf x_0,x_1,..., x_{\infty}, \forall\,\,} {\bf x}_i \in \mathbb{R}^2 \},
\end{equation}
with density $\lambda$. We also assume that the Voronoi cell $C({\bf x})$ which is defined as
\begin{equation}
C({\bf x}) = \{   {\bf\lVert x-y \rVert}^2 \le {\bf\lVert y-z \rVert}^2 ~~\forall ~~ {\bf x\in \textnormal{ $\Phi$}, z \in \textnormal{ $\Phi$} \backslash \{x\}, y\in } \mathbb{R}^2  \}
\end{equation}
may have one or more users. At a particular time instance, only a single user is served on a particular resource channel to avoid intra-cell interference. 
         
{\it  Large Scale Fading:} We assume that the large-scale fading model follows the ABG large-scale path-loss model \cite{sun2016propagation}, i.e., the path-loss can be written as 
\begin{equation}\label{eq:ABG_PATHLOSS}
L_i(h,r)_{dB} =  10\,\alpha_i\log_{10}({\sqrt{h^2+r^2}}) +\beta + 10\,\gamma_i \log_{10}({f}) +  \mathcal{X}_{\sigma_i}
\end{equation}
where $h$ is the vertical difference in height between the BS and the mobile user, $r$ is the horizontal distance between the mobile user and the BS, $\alpha_i$ is the path-loss exponent, $\mathcal{X}_{\sigma_i}$ is the shadow fading deviation in dB for the large-scale fluctuation and $\gamma$ and $\beta$ are the ABG path-loss parameters in dB as shown in Table \ref{table:pathloss_table}\footnote{For the sack of simplicity and tractability, we will neglect the effect of the log-normal shadowing parameter $\mathcal{X}_{\sigma_i}$ in this paper.}. The reason for adopting this model is that the model incorporates both NLoS and LoS propagation models and also provides a realistic and practical three-dimensional model that explicitly incorporates the height of the BS as a path-loss parameter. Moreover, this model is valid for UMi/UMa networks since it implicitly shows the valid ranges of the base station densities by giving the terrestrial distance range. In order to capture the actual effect of both the LoS/NLoS parts of the model, we need to know the probability that the user will have LoS connection with the BS at a certain height from the ground. We adopt the same model that is developed in \cite{al2014optimal}. Hence, the probability of the mobile user to be in LoS/NLoS with the associated BS can be written as
\begin{equation}
	\mathcal{P}_{L}(h,r) = \frac{1}{1+a\,{{\rm e}^{-b c\,\tan^{-1}\left(\frac{h}{r}\right)+b\,a}}},\,
	\mathcal{P}_{NL}(h,r) = 1 - \mathcal{P}_{L}(r),
\end{equation}
where $a$ and $b$ are environment-dependent constants with $c = 180/\pi$\footnote{From now on, we will write the subscript $L$ to refer to $LoS$ and use $NL$ to refer to $NLoS$.}.

{ \it Small Scale Fading:} It is assumed that large-scale path-loss is complemented with small-scale Rayleigh fading such that $\left|g\right|^2 \sim$ Exp(1) where $\left|g\right|^2 \sim$ is the channel gain between any arbitrary user and the BS. Also, it is assumed that the network is operating in an interference limited regime (i.e., performance of all links is limited by the co-channel interference and thermal noise at the receiver front-end is negligible). The assumption of the Rayleigh fading model is due to the simplicity of the analysis. This assumption yields the worst case performance and the analysis can be easily extended to a more generic Nakagami-m fading model. However, the effect of LoS and NLoS components is incorporated into the large-scale fading model giving by \eqref{eq:ABG_PATHLOSS}.

{\it Transmission Model and Channel Partitioning:} In this paper we assume that the mobile user is associated to the nearest BS (i.e., the BS which maximizes average received SNR) and transmitters on the same frequency are considered as co-channel interferers. The probability density function for the distance $R_1$ from the downlink user to the nearest BS assuming a HPPP can be written as
\begin{eqnarray}
f_{R_1}(r_1) = 2\pi r_1\lambda e ^{-\pi r_{1}^2\lambda}.
\end{eqnarray}
To decrease the level of the aggregate interference and increase network capacity, we assume that channel partitioning is applied (i.e., orthogonal frequency division multiple access (OFDMA) or time division multiple access (TDMA)). That is, the total BS channel bandwidth $W$ is shared in terms of time/frequency. Hence, the channel is partitioned in time/frequency into $N_s$ partitions (i.e., sub-carriers for OFDMA or time-slots for TDMA) and this number of partitions is assigned randomly into $N_a$ active users per cell. In addition, we will neglect the randomness of the number of active users in the cell and assume that $N_a$ is a fixed number (For more details on the distribution of $N_a$, you can refer to \cite{active_u_yu2013downlink}). Moreover, this kind of medium access scheme (i.e., channel partitioning) can be extended to any other medium access scheme (i.e., ALOHA, CSMA, CSMA-CA, etc.). 
       
\section{Performance analysis} \label{sec:analysis}
In this section, we study two main types of coverage performance: (i) the coverage probability and (ii) the rate coverage probability. For a complete performance analysis, we evaluate two main higher-order statistics of these metrics. Namely, we quantify the meta-distribution and the spatial capacity. These types of higher-order statistics provide a better insight into the two main types of cellular services. The first for a best effort network coverage probability and the second for a rate coverage of the network.

\subsection{Coverage Probability}
The coverage probability is defined as the probability that the SIR will be greater than a certain predefined value $\theta$. The average SIR at a downlink user located at the the origin can be quantified as
\begin{eqnarray} \label{eq:sir}
\sir=\underbrace{\frac{\left|g\right|^2~ {L}_{L}^{-1}(r_1) }{I_{\Phi^{}}} \mathcal{P}_{L}(r_1)}_{\sir_L} +\underbrace{\frac{\left|g\right|^2 {L}_{NL}^{-1}(r_1) }{I_{\Phi^{}}}\mathcal{P}_{NL}(r_1)}_{\sir_{NL}}.
\end{eqnarray}
Here, $\sir_L$ is SIR when there is a LoS link between the user and the BS, $\sir_{NL}$ is SIR when there is a NLoS link between the user and the BS and $I_{\Phi{}}$ is the aggregate interference from the co-channel transmitting BSs experienced by the mobile user and can be quantified as
\small
\begin{eqnarray} \label{eq:inter1}
	I_{\Phi^{}}^{} &=& I_{\Phi_{L}}^{} + I_{\Phi_{NL}}^{} \\ &=&\sum_{i\in\Phi_{}^{} \setminus\{0\}} \left|g_{}\right|^2 L_{L}(h,r_i) +\sum_{m\in\Phi_{}^{} \setminus\{0\}} \left|g_{}\right|^2  L_{NL}(h,r_m), \nonumber 
\end{eqnarray}
\normalsize
\normalfont
where $\Phi_{}$ is the set of all co-channel active BSs, $\Phi_{L}$ and $\Phi_{NL}$ are the set of all LoS and NLoS active base stations, respectively,  and $r_1$ is the horizontal distance from the mobile user to the nearest BS and $I_{\Phi_{L}}^{}$ and $I_{\Phi_{NL}}^{}$ are the aggregate interferences from the LoS and NLoS active base stations, respectively. Here, we assume that the channel power fading coefficients for the co-channel interferers, $\left|g_{}\right|^2$, are iid Rayleigh distributed random variables.

The coverage probability for any arbitrary mobile user can be evaluated as in the following theorem.
\begin{thm} \label{thm:thm1} {\bf (Coverage probability).} Coverage probability for any ergodic stationary PPP with a density $\lambda$ of BSs, $N_s$ channel partitions and $N_a$ active users per cell can be evaluated as
\begin{equation} \label{eq:coverage}\small
P_{\theta} =\int_{0}^{\infty} [\mathcal{P}_{L}(h,r_1)\textrm{$A(r_1,\theta)$} +\mathcal{P}_{NL}(h,r_1)\textrm{$B(r_1,\theta)$}] f_{R_1}(r_1) \dif r_1,
\end{equation}
\normalsize
where
\begin{IEEEeqnarray}{rCl} \label{eq:coverage_prob}
	A(r_1,\theta)= \exp \Big( - 2\pi \frac{\lambda N_a}{N_s}\int_{r_1}^{\infty}  1- \eta(s,r) \dif r\Big)|_{s = \frac{\theta}{L_{L}^{-1}(h,r_1)}  },\nonumber\\
	B(r_1,\theta) = \exp \Big( - 2\pi \nonumber \frac{\lambda N_a}{N_s}\int_{r_1}^{\infty}  1- \eta(s,r) \dif r\Big)|_{s = \frac{\theta}{L_{NL}^{-1}(h,r_1)}  }.
\end{IEEEeqnarray}
with $\eta(s,r)$ defined in Appendix \ref{app:app1}.
\begin{proof}
Please refer to Appendix \ref{app:app1} for proof.
\end{proof}
\end{thm}
In the results section, we will focus on the full load capacity of the network where the number of the active users $N_a$ requiring service on the same time in any cell  is equal to the number of channel partitions $N_s$. 
\subsection{Rate Coverage Probability}
The rate coverage is defined as the average probability at which the channel transmission rate will be greater than a certain level such that the rate QoS threshold $R_{o}$ will be achieved\footnote{Here, we denote the coverage rate probability as the short term coverage rate.}. The coverage rate for a certain threshold $R_{o}$ bits/s/Hz is defined as follows:
\begin{equation} 
P_{R_o} = \Pr\left[\frac{W}{N_s}{} \log_2 \left(1+\text{SIR}\right) \geq R_{o} \right] 
= \Pr[\text{SIR} \geq 2 ^{\frac{R_{o} N_s}{W}}-1].
\end{equation}
By intuition, the total channel bandwidth is divided into the $N_s$ number of channel partitions even for OFDMA or TDMA. For OFDMA it gives the effective channel bandwidth experienced that is associated to the user. For the TDMA scheme, it is the effective time utilization by the user, where $1/N_s$ is the effective normalized throughput of the maximum sum rate. Consequently, the coverage probability for the desired user at the origin can be quantified as in the next theorem.
\begin{thm} {\bf (Coverage rate probability).} The coverage rate for any ergodic stationary PPP with a density $\lambda$ of BSs, $N_s$ channel partitions and $N_a$ active users per cell can be evaluated as
\begin{equation} \label{eq:coverage_rate} \small
	P_{R_o} = \int_{0}^{\infty} [\mathcal{P}_{L}(h,r_1)\textrm{$A(r_1,R_o)$} +\mathcal{P}_{NL}(h,r_1)\textrm{$B(r_1,R_o)$}] f_{R_1}(r_1) \dif r_1,
\end{equation}
\normalsize
where
\begin{IEEEeqnarray}{rCl} \label{eq:coverage_rate_app}
	A(r_1,R_o)= \exp \Big( - 2\pi \frac{\lambda N_a}{N_s}\int_{r_1}^{\infty}  1- \eta(s,r) \dif r\Big)|_{s = \frac{2 ^{\frac{R_{o} N_s}{W}}-1}{L_{L}^{-1}(h,r_1)}  },\nonumber\\
	B(r_1,R_o)= \exp \Big( - 2\pi \nonumber \frac{\lambda N_a}{N_s}\int_{r_1}^{\infty}  1- \eta(s,r) \dif r\Big)|_{s = \frac{2 ^{\frac{R_{o} N_s}{W}}-1}{L_{NL}^{-1}(h,r_1)}},
\end{IEEEeqnarray}
with $\eta(s,r)$ defined in Appendix \ref{app:app1}.
	\begin{proof}
	The proof follows the same steps as for Theorem \ref{thm:thm1}.
	\end{proof}
\end{thm} 
For the derived formulas in \eqref{eq:coverage} and \eqref{eq:coverage_rate} and the rest of the paper, we will use the asterisk sub-script to refer to the optimal values of the parameters that maximize the chosen coverage function. In particular, we write 
\begin{equation} 
\{h^*,\lambda^*\}_{\theta} = \underset{h,\lambda}{\arg}~ \max  P_{\theta}
\end{equation}
for the coverage probability and
\begin{equation} 
\{h^*,\lambda^*\}_{R_o} = \underset{h,\lambda}{\arg}~ \max P_{R_o}
\end{equation}
for the rate coverage probability.
In the following section, we analyze meta-distribution for the considered cellular network.  
\subsection{Meta-distribution}

The coverage probability and rate coverage derived in  \eqref{eq:coverage} and \eqref{eq:coverage_rate} only provide average performance. Such an averaging does not provide an insight on the network level performance. From a network level perspective, the fraction of the users which can attain a certain desired level of coverage is important to quantify the quality-of-experience for the users. To this end, the authors in \cite{haenggi2016meta} introduced the meta-distribution of the coverage which is given as 
\begin{equation}
\bar{F}_{P_c}(x) \overset{\Delta}{=} \mathbb{P}^{!} \left(P_c > x\right).
\end{equation}
In other words, the meta-distribution is the complementary cumulative density function (CCDF) of the coverage probability. In this paper, we are interested in the coverage under the ABG path-loss model which implicitly accounts for LoS/NLoS propagation: 
\begin{equation}
\bar{F}_{P_{\theta}} (x) \overset{\Delta}{=} \mathbb{P}^{!}(P_{\theta}\ge x) \text{ and } \bar{F}_{P_{R_o}} (x) \overset{\Delta}{=} \mathbb{P}^{!}(P_{R_o}\ge x)
\end{equation}
where $\mathbb{P}^{!}$ is the Palm measure conditioning that the user is located at the origin. More clearly, the meta-distribution provides the probability that any arbitrary user in the network will achieve $P_{\theta}>x$, $P_{R_o}>x$ in $F_{P_{(.)}}\times 100\%$ of the time. The calculation of the meta-distribution is been made possible by the Gil-Pelaez theorem \cite{gil1951note} and can be quantified as
\begin{equation}\label{eq:gil1}
\bar{F}_{P_{\theta}} (x) =  \frac{1}{2} + \frac{1}{\pi}\int_{0}^{\infty} \frac{\text{Im} [e^{-jt\text{log}x }M_{jt}(\theta)]}{t} \dif t  ,
\end{equation}
and
\begin{equation}\label{eq:gil2}
\bar{F}_{P_{R_o}} (x) =  \frac{1}{2} + \frac{1}{\pi}\int_{0}^{\infty} \frac{\text{Im} [e^{-jt\text{log}x }M_{jt}(R_o)]}{t} \dif t,
\end{equation}
where $M_{jt}(\theta)$ and $M_{jt}(R_o)$ are the complex ${jt}^{th}$ moments of $P_{\theta}$ and $P_{R_o}$, respectively and Im[.] is the imaginary part symbol with $j = \sqrt{-1}$. In order to find exact expressions of the meta-distribution, we need to find the expressions for $M_{m}(\theta)$ and $M_{m}(R_o)$ which are the real $m^{th}$ moments. This can be quantified as in the following theorem.
\begin{thm} \label{thm:thm3}{\bf (Moments)} The moments $M_{m}(\theta)$ and $M_{m}(R_o)$ for any ergodic stationary PPP with density $\lambda$, $N_s$ channel partitions and $N_a$ active users per cell can be evaluated as
	
	\small
 	\begin{IEEEeqnarray}{rCl} \label{eq:mtheta} 
 		M_{m}(\theta) &=& \int_{0}^{\infty} [\mathcal{P}_{L}(h,r_1)\textrm{$A_m(r_1,\theta)$}  \nonumber \\ && ~~ +\mathcal{P}_{NL}(h,r_1)\textrm{$B_m(r_1,\theta)$}] f_{R_1}(r_1) \dif r_1, \\
 			\label{eq:m_r} 
 	M_{m}(R_o) &=& \int_{0}^{\infty} [\mathcal{P}_{L}(h,r_1)\textrm{$A_m(r_1,R_o)$} \nonumber \\ && ~~+\mathcal{P}_{NL}(h,r_1)\textrm{$B_m(r_1,R_o) $}] f_{R_1}(r_1) \dif r_1,
 	\end{IEEEeqnarray}
 \normalsize
 	where
 	\small
 	\begin{IEEEeqnarray}{rCl} \label{eq:moment_App}
 		A_m(r_1,\theta)= \exp \Big( - 2\pi \frac{\lambda N_a}{N_s}\int_{r_1}^{\infty}  1- \eta_m(s,r) \dif r\Big)|_{s = \frac{\theta}{L_{L}^{-1}(h,r_1)}},\nonumber\\
 		B_m(r_1,\theta) = \exp \Big( - 2\pi \nonumber \frac{\lambda N_a}{N_s}\int_{r_1}^{\infty}  1- \eta_m(s,r) \dif r\Big)|_{s = \frac{\theta}{L_{NL}^{-1}(h,r_1)}  }, \\
 		A_m(r_1,R_o)= \exp \Big( - 2\pi \frac{\lambda N_a}{N_s}\int_{r_1}^{\infty}  1-\eta_m (s,r) \dif r\Big)|_{s = \frac{2 ^{\frac{R_{o} N_s}{W}}-1}{L_{L}^{-1}(h,r_1)}  },\nonumber\\
 		B_m(r_1,R_o) = \exp \Big( - 2\pi \nonumber \frac{\lambda N_a}{N_s}\int_{r_1}^{\infty}  1- \eta_m(s,r) \dif r\Big)|_{s = \frac{2 ^{\frac{R_{o} N_s}{W}}-1}{L_{NL}^{-1}(h,r_1)} },
 	\end{IEEEeqnarray}
 with $\eta_m(s,r)$ defined in Appendix \ref{app:app2}.
 	\begin{proof}
 		Please refer to Appendix \ref{app:app2} for proof.
 	\end{proof}
 \end{thm}
\normalsize
Here, substituting \eqref{eq:mtheta} and \eqref{eq:m_r} into \eqref{eq:gil1} and \eqref{eq:gil2} gives exact expressions of the meta-distributions. Unfortunately, this way of evaluating the meta-distribution is intractable and requires a long time for the numerical integrations. To make this more tractable, an excellent precise approximation of the Gil-Pelaez theorem can be obtained by utilizing Mnatsakanov's theorem \cite{dist_rec}. Using Mnatsakanov's theorem, we will be able to recover the distribution of any arbitrary random variable, conditioned on the requirement that any real integer's $m^{th}$ moment is defined. Hence, the meta-distribution can be given in approximate value as
\begin{eqnarray} \label{eq:approx1}
\bar{F}_{P_{\theta}} (x) \approxeq \sum_{k=0}^{[\mu x]}\sum_{j=k}^{\mu} \binom{\mu}{j}\binom{j}{k}(-1)^{j-k} M_{m}(\theta),\\\label{eq:approx2}
\bar{F}_{P_{R_o}} (x) \approxeq \sum_{k=0}^{[\mu x]}\sum_{j=k}^{\mu} \binom{\mu}{j}\binom{j}{k}(-1)^{j-k} M_{m}(R_o),
\end{eqnarray}
where $\mu$ is an arbitrary integer such that the larger it is the more accurate is the approximation. We choose this approximation due to its fast convergence to the exact solution which is evaluated by the Gil-Pelaez theorem which requires integrations of complex numbers. As we will show in the results section, the first $25$ moments will be sufficient to precisely recover the distribution. From the $m^{th}$ moments in \eqref{eq:mtheta} and \eqref{eq:m_r}, we evaluate the second cumulants (variances) for both coverage probability and rate coverage probability as
\begin{eqnarray}
&&var(\theta) = 	M_{2}(\theta) -M_{2}^2(\theta), \\
&&var(R_o)    = 	M_{2}(R_o) -M_{2}^2(R_o).
\end{eqnarray}
These variances provide more insight on the spread of the coverage values over all the users for a certain desired threshold and its deviation from the average value. Intuitively, the less the variance, the better is the fairness between the network users in terms of coverage.

\subsection{Spatial Coverage Capacity and Spatial Rate Capacity}
In order to answer the question \quotes{What is the maximum density of the concurrent active users that satisfy a certain predefined coverage reliability?} we derive expressions for the spatial coverage capacity and spatial rate capacity. These metrics provide fine grained characteristics of the cellular network and network level service quality.
 
\begin{definition} {\bf (Spatial coverage capacity).} The spatial coverage rate for any ergodic stationary PPP with density $\lambda$ of BSs, $N_s$ channel partitions and $N_a$ active users per cell is defined as the maximum effective density of users that have SIR values greater than the QoS threshold $\theta$ with probability at least $P_{\theta}=x$ and can be evaluated as
\begin{equation} \label{eq:scc}
SCC(x,\theta,N_s) \overset{\Delta}{=} N_a \lambda \bar{F}_{P_{\theta}} (x),
\end{equation}
and the optimal operating point for network full capacity is defined as
\begin{equation} \label{eq:opt1}
\{\lambda^*,h^*,N_s^*\} \overset{\Delta}{=}\underset{\lambda,h,N_s}{\arg} SCC(x,\theta,N_s).
\end{equation}
\end{definition} 
Here, we can use simple two dimensional search algorithms to find this optimal operating point.                                         
\begin{definition} {\bf (Spatial rate capacity).} The spatial rate capacity for any ergodic stationary PPP with density $\lambda$ of BSs, $N_s$ channel partitions and $N_a$ active users per cell is defined as the maximum effective density of users that have channel rate values greater than the QoS threshold $R_o$ with probability at least $P_{R_o}=x$ and can be evaluated as
	\begin{equation} \label{eq:src}
	SRC(x,R_o,N_s) \overset{\Delta}{=} N_a \lambda \bar{F}_{P_{R_o}} (x),
	\end{equation}
	and the optimal operating point for network full capacity is defined by
	\begin{equation}\label{eq:opt2}
	\{\lambda^*,h^*,N_s^*\} \overset{\Delta}{=}\underset{\lambda,h,N_s}{\arg} SRC(x,R_o,N_s).
	\end{equation}
\end{definition}    
Using the accurate approximation we introduced in \eqref{eq:approx1} and \eqref{eq:approx2}, we can easily find the solution for \eqref{eq:opt1} and \eqref{eq:opt2} without the need for applying the Gil-Pelaez theorem.

From the above analysis of the network performance metrics, we can build a comprehensive framework for analysing network level performance and capture the individual and spatial performance in a fine-grained strategy instead of only looking to the standard average coverage probability metrics.

\section{Results discussion} \label{discussion}
In this section, we present numerical results for the given evaluated metrics. We will assume an urban environment with the parameters $a =9.6$, $b =0.28$, $f = 2$ GHz carrier frequency and the BS total available bandwidth $W = 20$ MHz. Also, as described in the previous sections, we consider Rayleigh flat wireless fading channels.
\subsection{Impact of Network Densification on Optimal Height and Optimal Average Coverage Probability}
\begin{figure}[!t] 
	\begin{center}
		\scalebox{0.40}{\includegraphics{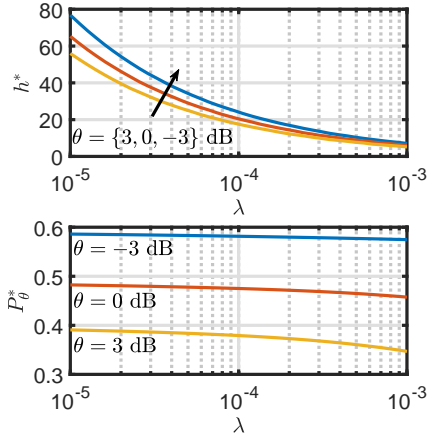}}
		\caption{Optimal values for the BS height (top) and the corresponding coverage probability (bottom) using the ABG-UMi path-loss model for different values of coverage SIR threshold $\theta$.}
		\label{fig:optimalh}
	\end{center}
\end{figure}

In \figref{fig:optimalh}, we show the optimal BS height and the corresponding optimal values of the coverage probability against BS densities of the network for different values of the SIR $\theta$ threshold. In this figure, we clearly observe that the optimal BS height changes as we vary the BSs density. That is, with the adopted LoS/NLoS model of large-scale fading, there is an optimal operational height at which the network operator will gain nearly the same coverage probability for any chosen base station density. Moreover, the chosen height of the BSs does explicitly depend on the SIR $\theta$ threshold which is pre-defined as the QoS metric. This also applies to the rate coverage probability ($P_{R_o}$) which must be optimized in parallel with the coverage probability.
\subsection{Optimal Parameter Selection Considering the Variance of the Received SIR}
\begin{figure*}[t]
	\centering
	\begin{subfigure}[b]{0.23\textwidth}
		\includegraphics[width=1.1\textwidth,height=1.1\textwidth]{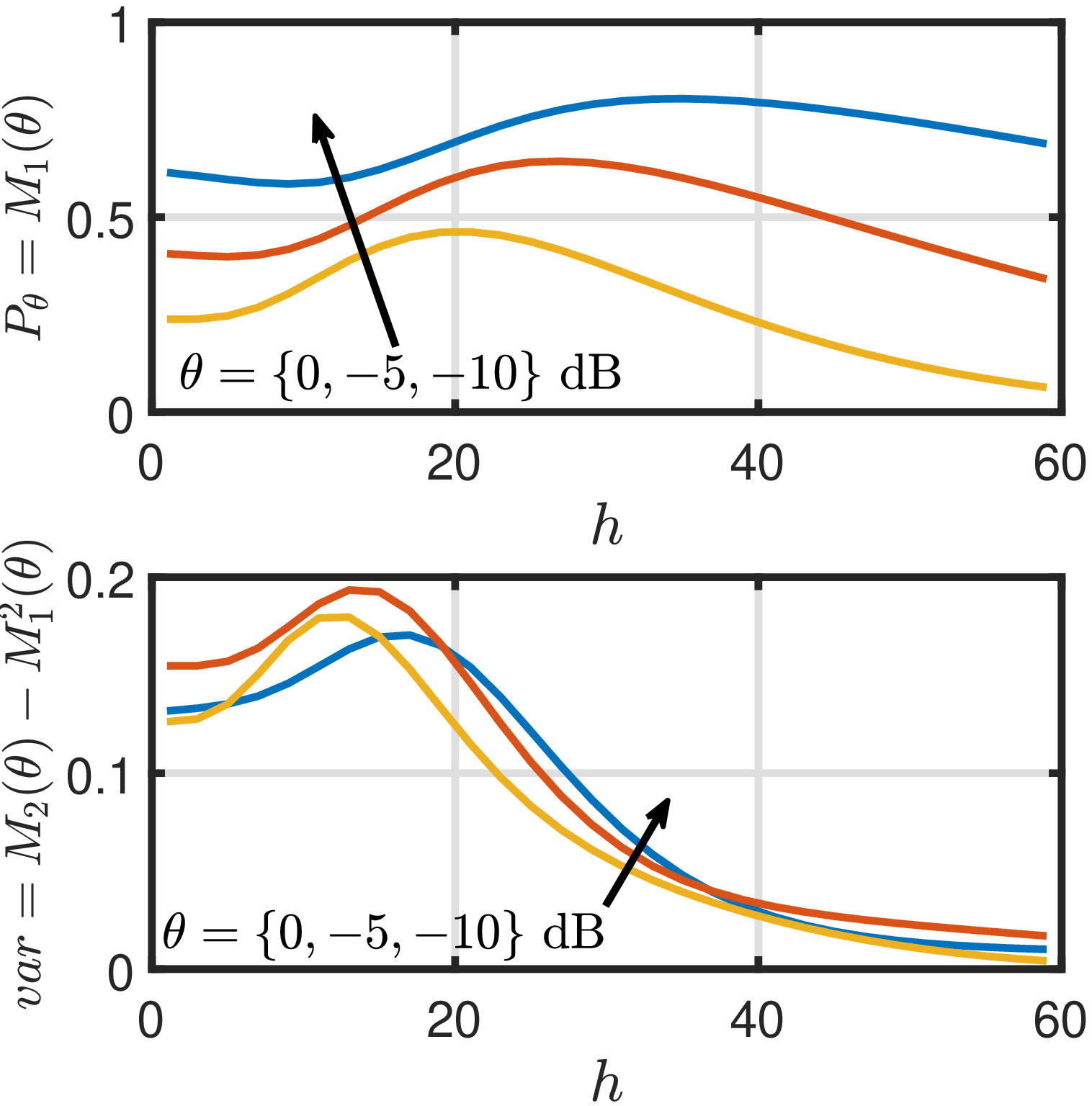}
		\caption{$\lambda  = 1\times 10 ^{-4}$.}
		\label{fig:COV_VAR_SINGLE}
	\end{subfigure}
	\quad 
	\begin{subfigure}[b]{0.23\textwidth}
		\includegraphics[width=1.1\textwidth,height=1.1\textwidth]{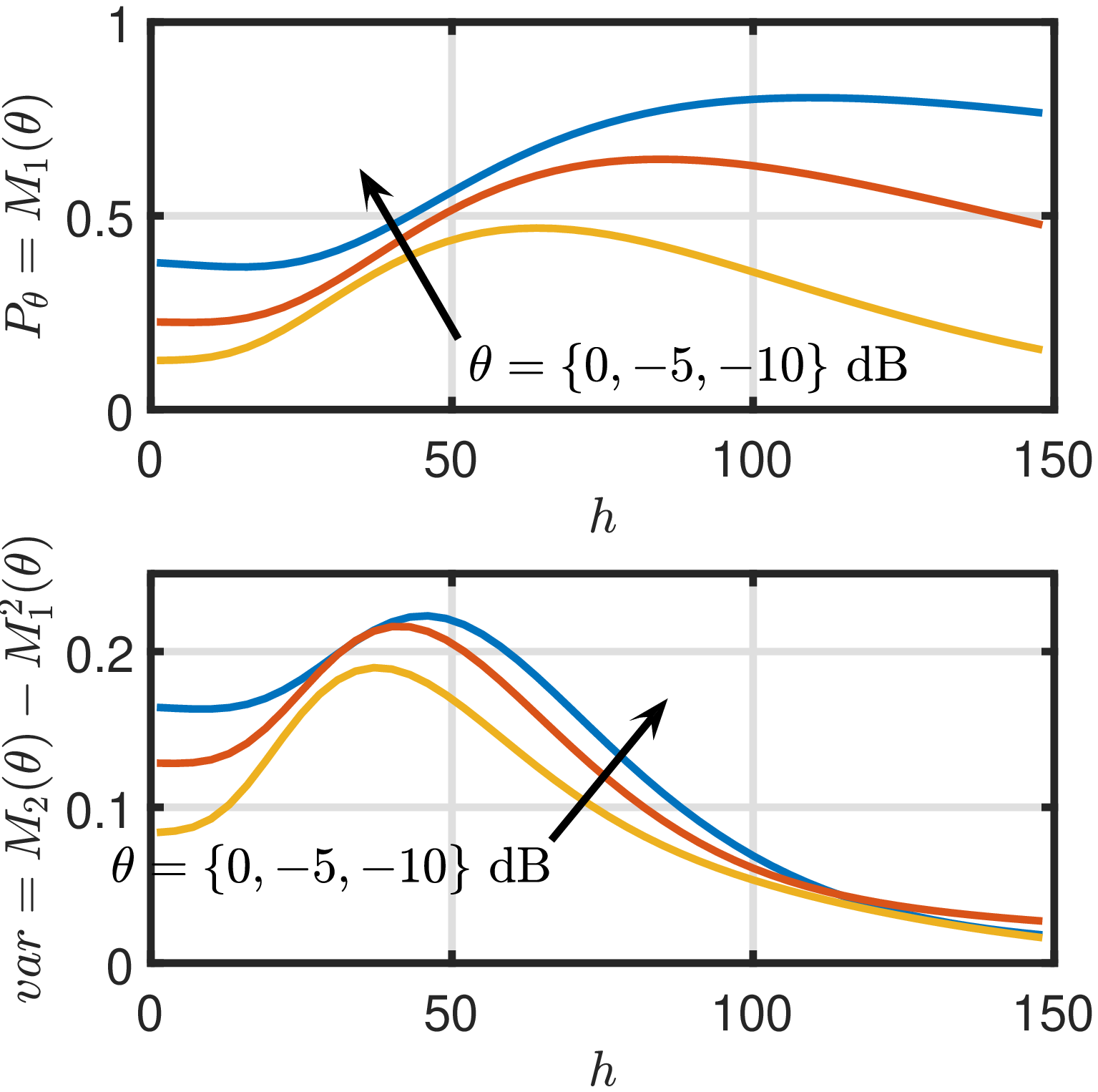}
		\caption{$\lambda  = 1\times 10 ^{-5}$.}
		\label{fig:COV_VAR_SINGLE_Eve5}
	\end{subfigure}
	\quad 
	\begin{subfigure}[b]{0.23\textwidth}
		\includegraphics[width=1.1\textwidth,height=1.1\textwidth]{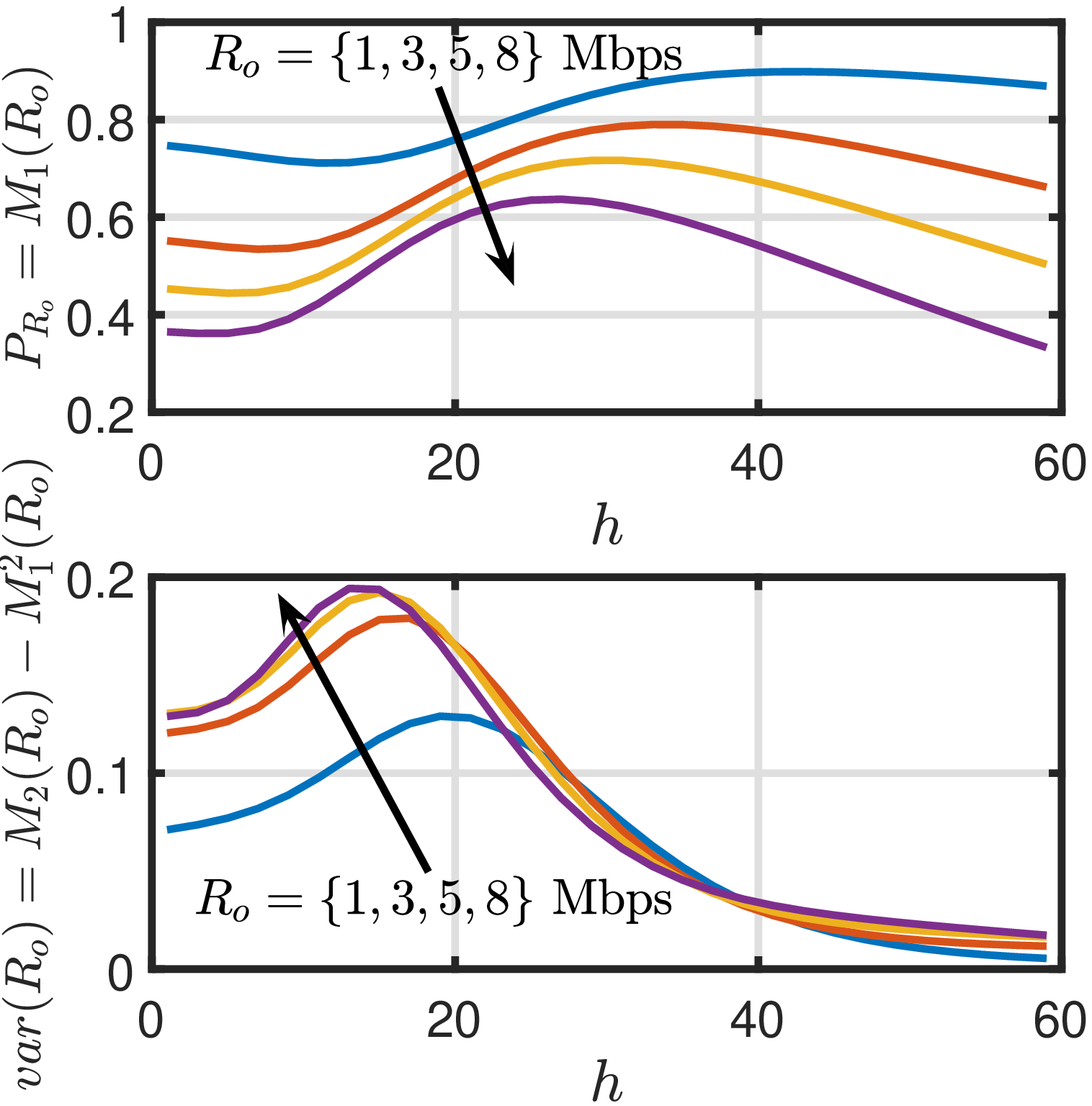}
		\caption{$\lambda  = 1\times 10 ^{-4}$.}
		\label{fig:RATE_COV_SINGLE_E_ve4}
	\end{subfigure}
	\quad 
	\begin{subfigure}[b]{0.23\textwidth}
		\includegraphics[width=1.1\textwidth,height=1.1\textwidth]{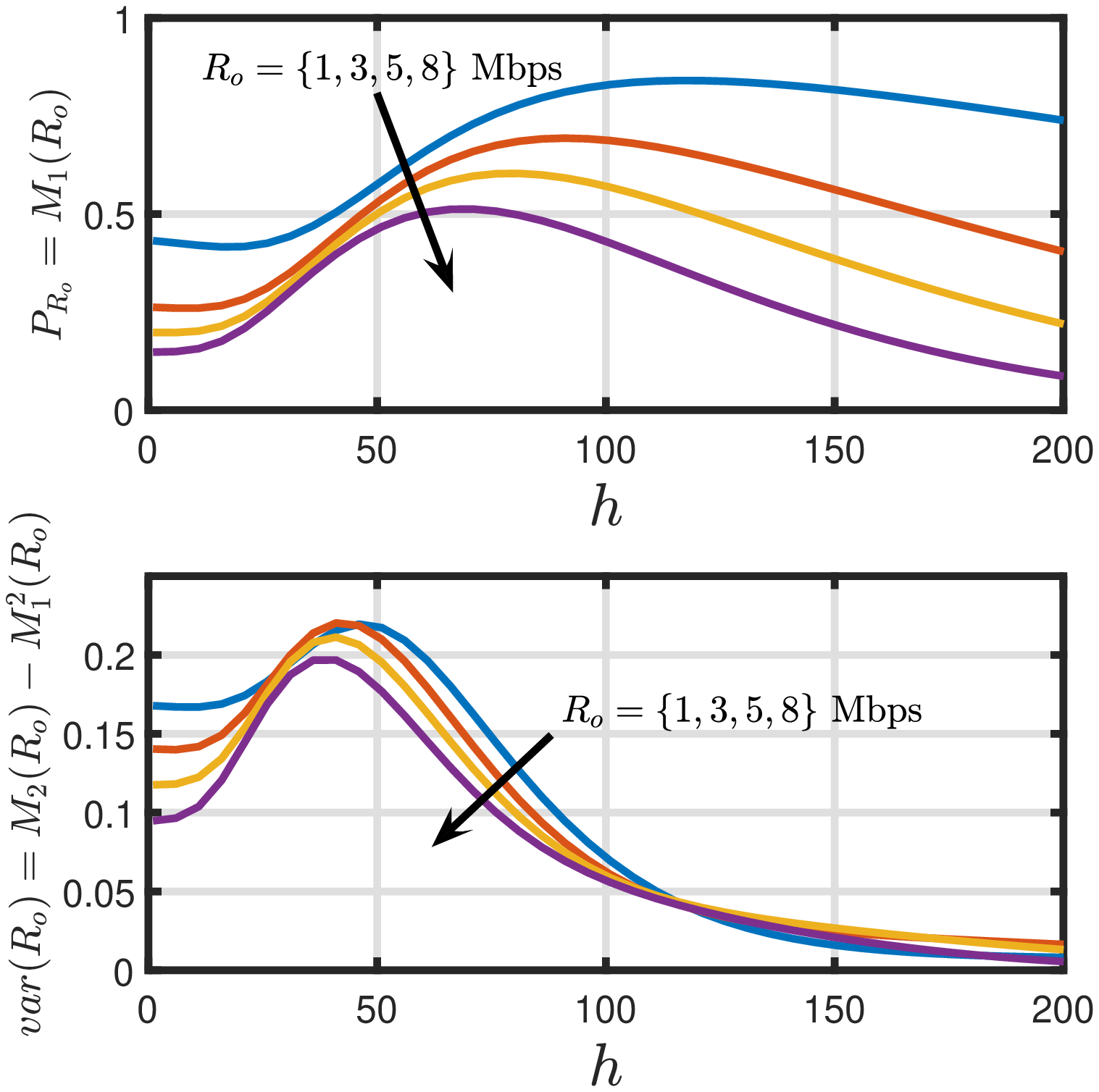}
		\caption{$\lambda  = 1\times 10 ^{-5}$.}
		\label{fig:RATE_COV_SINGLE_E_ve5}
	\end{subfigure}
	\caption{ (a)+(b) Coverage probability and coverage variance. (c)+(d) Coverage rate probability and coverage rate variance. All for the UMi large-scale fading model, $W = 20$ MHz and $N_a = N_s = 1$. }
	\label{fig:variances}
	\vspace*{4pt}
\end{figure*}

\figref{fig:variances} shows both the first and second cumulants of the coverage probabilities (i.e., the mean as the coverage probability/rate coverage probability and the variance as a deviation measure). The main motivation for studying the variance is that it is considered as the most important measure of fairness between the users. The most interesting point here is that the slope of the variance curves is much steeper for BS heights which are greater than the optimal height $h^*$ that maximizes the first moment. That is, increasing $h$ beyond $h^*$ will slightly decrease $P_{\theta}$ and $P_{R_o}$, but decreases the variances more rapidly. Hence, the network operator may choose to sacrifice the optimal values of $P_{\theta}$ and $P_{R_o}$ to gain more fairness for the users. For example, in \figref{fig:variances}.(c) the height that maximizes $P_{R_o}$ for $R_o = 8$ Mbps is around $25$ meters which corresponds to a variance of $0.13$. However, increasing the height of the base station by $5$ meters will result in a slight decrease in $P_{R_o}$ by $.05$ and will also decrease the variance to $0.07$ which is approximately half of $0.13$. That is, an additional $70\%$ more users will gain the same optimal value of $P_{R_o}$. More clearly, this will increase the user's fairness over the entire network. However, this behaviour needs to be considered carefully due to the large number of parameters involved in the network radio access planning.

\subsection{Evaluation of Meta-distribution and Optimal Bandwidth Partitioning}
\begin{figure}[!t] 
	\begin{center}
		\scalebox{0.45}{\includegraphics{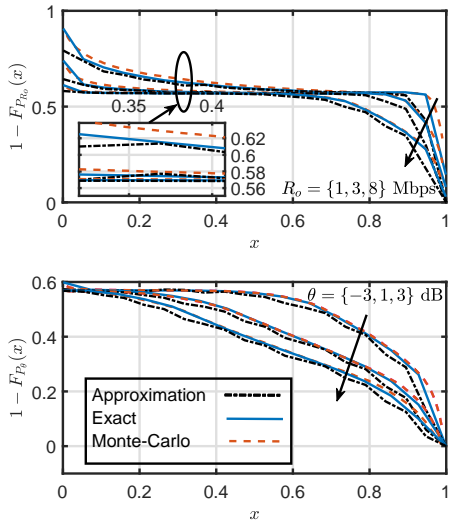}}
		\caption{Meta-distribution of coverage rate probability (top) and meta-distribution for coverage probability (bottom). All for the UMi large-scale fading model, $W = 20$ MHz and $N_a=N_s = 1$.}
		\label{fig:meta_validiations}
	\end{center}
\end{figure}

\begin{figure}[!t] 
	\centering
	\scalebox{0.5}{\includegraphics{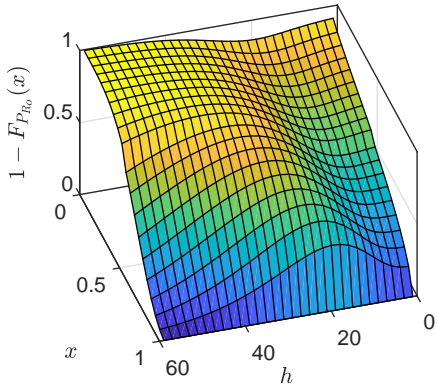}}
	\caption{Meta-distribution for coverage rate for single tier network with $\lambda = 1\times 10^{-4}$, $R_o = 3$ Mbps and $N_s=N_a = 1$.}
	\label{fig:d3d_plot}
\end{figure}

\figref{fig:meta_validiations} shows a comparison between the exact solution (see \eqref{eq:gil1} and \eqref{eq:gil1}), approximation (see \eqref{eq:approx1} and \eqref{eq:approx2}) and Monte-Carlo simulations for the meta-distribution. As seen from the plot, the approximation that we used matches the exact solution and the Monte-Carlo simulation. The three bunches of curves are for different values of $\theta$ and $R_o$, but for the same system parameters and BSs density. As expected, the corresponding values of the meta-distribution are different for the three curves. This means that $P_{\theta}$ and $P_{R_o}$ do not provide sufficient information about the network performance. For example, the values for meta-distribution $\bar{F}_{P_{\theta=-3 \text{dB}}} (0.8) = 0.40$ while $\bar{F}_{P_{\theta=0 \text{dB}}} (0.8) = 0.27$. That is, $13\%$ less users at $\theta=0$ dB QoS will not achieve the $0.8$ coverage probability as compared to the value at $\theta=-3$ dB. Another interesting point is that, with the optimal height deployment, the meta-distribution is less likely to have zero values and the curves are more likely to be flat. In some papers, the meta-distribution is approximated (using the first two or three moments) by the beta-distribution and the generalized beta-distribution \cite{deng2017fine}. But, this is not valid for our model and so dramatically fails. This is due to the fact that the optimal height deployment is more favourable for the environmental conditions and results in more LoS links, especially for the nearest neighbour association which flattens the curve and decreases the variance (i.e., more fairness between users - see \cite{deng2017fine} for more details). Hence, the two or three parameters distribution mapping like the beta-distribution and the generalized beta-distribution is not sufficient and this is why we have used the Mnatsakanov's theorem as an approximation. Finally, \figref{fig:d3d_plot} shows the effect of changing the BSs height on the meta-distribution. As shown in this figure, for any arbitrarily chosen value of $x$, there is an optimal height at which the meta-distribution will be maximized. This is valid for both $\bar{F}_{P_{\theta}} (x)$ and $\bar{F}_{P_{R_o}} (x)$. \\

\begin{figure}[!t]
	\centering
	\begin{subfigure}[b]{0.23\textwidth}
	\scalebox{0.23}{\includegraphics[]{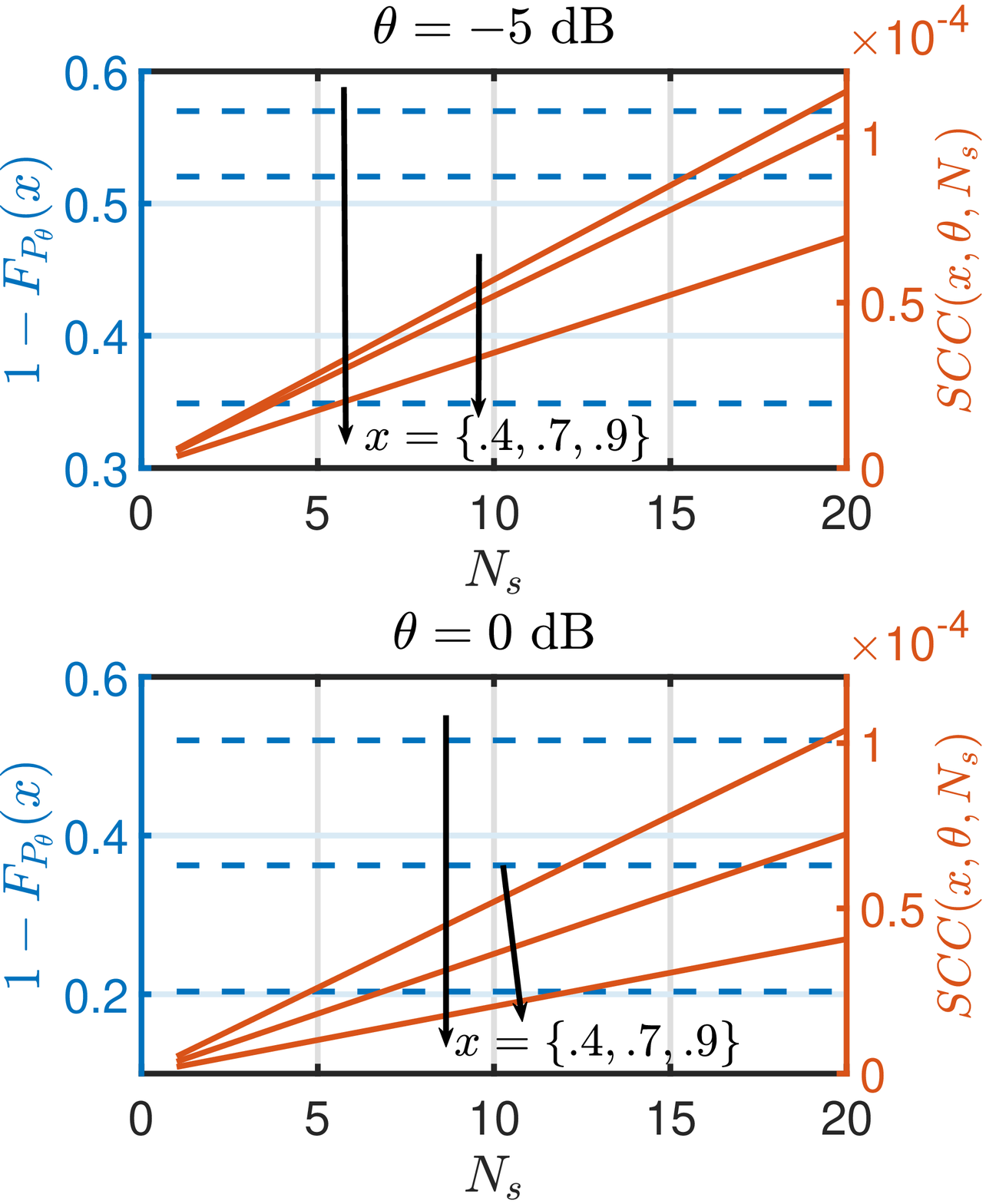}}
	\label{fig:fmeta_SCC}
	\subcaption{}
	\end{subfigure}
	\quad 
	\begin{subfigure}[b]{0.23\textwidth}
		\scalebox{0.23}{\includegraphics[]{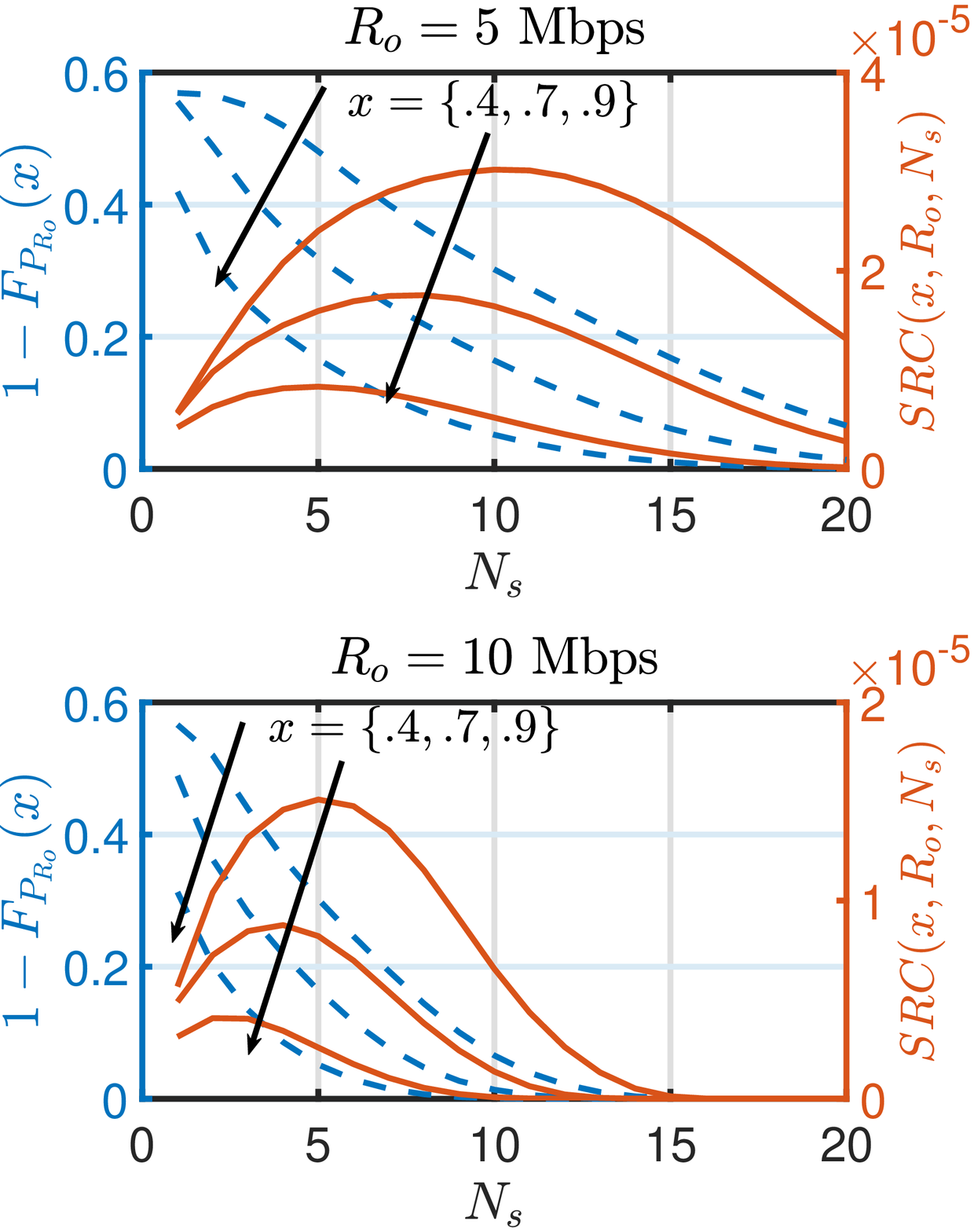}}
	\label{fig:fmeta_SRC}
	\subcaption{}
	\end{subfigure}
	\caption{ (a) Meta-distribution (left axis- see \eqref{eq:approx1}) and SCC (right axis - see \eqref{eq:scc}) against the number of channel partitions. (b) Meta-distribution (left axis- see \eqref{eq:approx1}) and SRC (right axis - see \eqref{eq:scc}) against the number of channel partitions. All for UMi large-scale fading model, $W = 20$ MHz and for full load cell $N_a=N_s$. }
	\label{fig:SPATIALS}
	\vspace*{4pt}
\end{figure}

\figref{fig:SPATIALS} shows the meta-distributions ($\bar{F}_{P_{\theta}} (x)$ and $\bar{F}_{P_{R_o}} (x)$) and the spatial capacities $SCC(x,\theta,N_s=N_a)$ and $SRC(x,R_o,N_s=N_a)$. An interesting point about the full load SRC is that for any arbitrarily chosen value of reliability $x$, there is an optimal number of channel partitions $N_s$ that maximizes the SRC. This optimal $N_s$ varies with the desired rate threshold $R_o$. For example, for $x=0.4$ and $R_o = 5$ Mbps, the optimal number of channel partitions for the full load capacity is $N_s = N_a = 10$  and the density of users who achieve $R_o$ is $3\times10^{-5}$ while for $N_s =18$ there are $33\%$ less users who achieve the same $R_o$. As clearly shown in the figure, this optimal number of partitions selections is only valid when studying the $SRC$ and it is not valid for the $SCC$ where there are no optimal values for $N_s$. This is due to the fact that the effective rate is dependent on the number of channel partitions and is a logarithmic function of the SIR while the standard coverage probability is not.


%
\section{Conclusion} \label{sec:conclusion}
In this paper, we highlighted some important aspects of the design for the radio access of the ultra-dense and traditional cellular networks. We gave expressions to quantify the coverage probability and coverage rate probability. For the sake of better characterization of the network performance and we evaluated the higher-order moments for both the coverage probability and coverage rate probability. As a result of the higher-order moments, we quantified the meta-distribution to characterize the users' fairness experience using the exact solution of Gil-Pelaez and also Mnatsakanov's theorem for an accurate approximation. Using the evaluated performance metrics, we studied the impact of changing any of the main system model parameters on the overall performance of the network. As the main result, we showed that both the partitioning factor of the channel and the BS height play very important roles in optimizing the network performance. Finally, for a future extension, we will study the same performance metrics for a multi-tier, user-centric heterogeneous network.

\appendices
\section{Proof of Theorem \ref{thm:thm1}} \label{app:app1}
The coverage probability is given by
\small
\begin{IEEEeqnarray}{rCl}
	P_{\theta} &=& \Pr \left[ \text{SIR} \ge \theta \right], \nonumber \\
	&=& \mathbb{E}_{r_1} \Big[\mathcal{P}_{L}(h,r_1)\underbrace{{\Pr\left[  \frac{\left|g\right|^2~ {L}_{L}^{-1}(r_1) }{I_{\Phi^{}}} \ge \theta \right]}}_\textrm{$A(r_1,\theta)$} \nonumber \\&&~~~~~~~~~~~~~~~~+\mathcal{P}_{NL}(h,r_1)\underbrace{{\Pr\left[ \frac{\left|g\right|^2~ {L}_{NL}^{-1}(r_1) }{I_{\Phi^{}}} \ge \theta \right]}}_\textrm{$B(r_1,\theta)$}\Big],\,\,\,~~
\end{IEEEeqnarray}
\normalsize
where
\scriptsize
\begin{eqnarray}
\!\!\!\!\!\!\!\!&& A(r_1,\theta) =\mathbb{E}_{I_{\Phi_{L}},I_{\Phi_{NL}}}[\exp(-sI_{\Phi})], \nonumber\\
\!\!\!\!\!\!\!\!	&&\overset{}{=}  \mathbb{E}_{\left|g\right|^2,I_{\Phi_{L}},I_{\Phi_{NL}}}[\prod_{ \underset{i \in \Phi_{L}}{m \in \Phi_{NL}} }\exp\left(-s\left|g\right|^2 (L_{L}^{-1}(h,r_i) +L_{NL}^{-1}(h,r_m))\right)],\nonumber\\
\!\!\!\!\!\!\!\!	&&\overset{(a)}{=}\mathbb{E}_{I_{\Phi_{L}},I_{\Phi_{NL}}}[\prod_{ \underset{i \in \Phi_{L}}{m \in \Phi_{NL}} } \underbrace{\frac{\mathcal{P}_{L}(r_i)}{1+sL_{L}^{-1}(h,r)} + \frac{\mathcal{P}_{NL}(r_m)}{1+sL_{NL}^{-1}(h,r_m)}}_{\eta(s,r)}],\nonumber\\
\!\!\!\!\!\!\!\!	&&\overset{(b)}{=} \exp \Big( - 2\pi \frac{\lambda N_a}{N_s}\int_{r_1}^{\infty}  1- \eta(s,r) \dif r\Big),
\mkern-18mu
\end{eqnarray}
\normalfont
\normalsize
 with $(a)$ is obtained by taking the expectation over the Rayleigh fading channel coefficient $\left|g\right|^2$, $(b)$ is obtained by applying the probability generating functional (PGFL) of the PPP, $s = \theta/L_{L}^{-1}(h,r_1)$ and $B((r_1,\theta)$ can be obtained in the same way as $A(r_1,\theta)$ by substituting $s$ in $A((r_1,\theta)$ by $s = \theta/L_{NL}^{-1}(h,r_1)$. For the rate coverage $P_{R_o}$, with $R_o$ we only substitute any $\theta$ by $2 ^{\frac{R_{o} N_s}{W}}-1$.
\section{Proof of Theorem \ref{thm:thm3}} \label{app:app2} 
The $m^{th}$ moment $M_{m}(\theta)$ and $M_{m}(R_o)$ can be evaluated as
\begin{IEEEeqnarray}{rCl}
	M_{m}(.) &=& \mathbb{E} [P_i^m],\,\,\, i \in \{{\theta},{R_o}\}\nonumber\\
	&=& \mathbb{E}_{r_1} \left[\mathcal{P}_{L}(h,r_1)\textrm{$A_m(r_1,\theta)$} \nonumber +\mathcal{P}_{NL}(h,r_1)\textrm{$B_m(r_1,\theta)$} \right],
\end{IEEEeqnarray}
\normalsize
where
\scriptsize
\begin{eqnarray}
\!\!\!\!\!\!\!\!&&A_m(r_1,\theta)=\mathbb{E}_{I_{\Phi_{L}},I_{\Phi_{NL}}}[\exp(-sI_{\Phi})^m], \nonumber\\
\!\!\!\!&&\overset{(a)}{=}  \mathbb{E}_{I_{\Phi_{L}},I_{\Phi_{NL}}}[\prod_{ \underset{i \in \Phi_{L}}{k \in \Phi_{NL}} } \underbrace{\frac{\mathcal{P}_{L}(r_i)}{(1+sL_{L}^{-1}(h,r))^m} + \frac{\mathcal{P}_{NL}(r_k)}{(1+sL_{NL}^{-1}(h,r_m))^m}}_{\eta_m(s,r)}],\nonumber\\
\!\!\!\!&&\overset{(b)}{=} \exp \Big( - 2\pi \frac{\lambda N_a}{N_s}\int_{r_1}^{\infty}  1- \eta_m(s,r) \dif r\Big),
\
\end{eqnarray}
\normalsize
 with $(a)$ is obtained by taking the expectation over the Rayleigh fading channel coefficient $\left|g\right|^2$, $(b)$ is obtained by applying the PGFL, $s = \theta/L_{L}^{-1}(h,r_1)$ and $B_m(r_1,\theta)$ can be obtained in the same way as $A_m(r_1,\theta)$ by substituting $s$ in $A_m(r_1,\theta)$ by $s = \theta/L_{NL}^{-1}(h,r_1)$ and $\theta$ by $R_o$. For the rate coverage $P_{R_o}$ with $R_o$, we only substitute any $\theta$ by $2 ^{\frac{R_{o} N_s}{W}}-1$.
 \normalsize
\bibliographystyle{ieeetr}
\bibliography{all}

\begin{thebibliography}{10}

\bibitem{andrews2011tractable}
J.~G. Andrews, F.~Baccelli, and R.~K. Ganti, ``A tractable approach to coverage
  and rate in cellular networks,'' {\em IEEE Transactions on communications},
  vol.~59, no.~11, pp.~3122--3134, 2011.

\bibitem{wang2018sir}
Y.~Wang, M.~Haenggi, and Z.~Tan, ``Sir meta distribution of k-tier downlink
  heterogeneous cellular networks with cell range expansion,'' {\em arXiv
  preprint arXiv:1803.00182}, 2018.

\bibitem{elsawy2017modeling}
H.~ElSawy, A.~Sultan-Salem, M.-S. Alouini, and M.~Z. Win, ``Modeling and
  analysis of cellular networks using stochastic geometry: A tutorial,'' {\em
  IEEE Communications Surveys \& Tutorials}, vol.~19, no.~1, pp.~167--203,
  2017.

\bibitem{kalamkar2017spatial}
S.~S. Kalamkar and M.~Haenggi, ``The spatial outage capacity of wireless
  networks,'' {\em IEEE Transactions on Wireless Communications}, 2018.

\bibitem{multi_slope_4_comprehensive}
M.~Ding, P.~Wang, D.~L{\'o}pez-P{\'e}rez, G.~Mao, and Z.~Lin, ``Performance
  impact of los and nlos transmissions in dense cellular networks,'' {\em IEEE
  Transactions on Wireless Communications}, vol.~15, no.~3, pp.~2365--2380,
  2016.

\bibitem{haenggi2016meta}
M.~Haenggi, ``The meta distribution of the sir in poisson bipolar and cellular
  networks,'' {\em IEEE Transactions on Wireless Communications}, vol.~15,
  no.~4, pp.~2577--2589, 2016.

\bibitem{ding2016please}
M.~Ding and D.~L. P{\'e}rez, ``Please lower small cell antenna heights in 5g,''
  in {\em Global Communications Conference (GLOBECOM), 2016 IEEE}, pp.~1--6,
  IEEE, 2016.

\bibitem{haenggi_70_pound}
M.~Haenggi, {\em Stochastic geometry for wireless networks}.
\newblock Cambridge University Press, 2012.

\bibitem{wang2017meta}
Y.~Wang, M.~Haenggi, and Z.~Tan, ``The meta distribution of the sir for
  cellular networks with power control,'' {\em IEEE Transactions on
  Communications}, 2017.

\bibitem{hayajneh2017performance}
A.~M. Hayajneh, S.~A.~R. Zaidi, D.~C. McLernon, and M.~Ghogho, ``Performance
  analysis of uav enabled disaster recovery network: A stochastic geometric
  framework based on matern cluster processes,'' in {\em Third Intelligent
  Signal Processing Conference Proceedings (ISP 2017)}, Institution of
  Engineering and Technology, 2017.

\bibitem{hayajneh2018performance}
A.~M. Hayajneh, S.~A.~R. Zaidi, D.~C. McLernon, M.~Di~Renzo, and M.~Ghogho,
  ``Performance analysis of uav enabled disaster recovery networks: A
  stochastic geometric framework based on cluster processes,'' {\em IEEE
  Access}, vol.~6, pp.~26215--26230, 2018.

\bibitem{deng2017fine}
N.~Deng and M.~Haenggi, ``A fine-grained analysis of millimeter-wave
  device-to-device networks,'' {\em IEEE Transactions on Communications},
  vol.~65, no.~11, pp.~4940--4954, 2017.

\bibitem{win2006error}
M.~Z. Win, P.~C. Pinto, A.~Giorgetti, M.~Chiani, and L.~A. Shepp, ``Error
  performance of ultrawideband systems in a poisson field of narrowband
  interferers,'' in {\em Spread Spectrum Techniques and Applications, 2006 IEEE
  Ninth International Symposium on}, pp.~410--416, IEEE, 2006.

\bibitem{atzeni2018downlink}
I.~Atzeni, J.~Arnau, and M.~Kountouris, ``Downlink cellular network analysis
  with los/nlos propagation and elevated base stations,'' {\em IEEE
  Transactions on Wireless Communications}, vol.~17, no.~1, pp.~142--156, 2018.

\bibitem{multi_slop}
X.~Zhang and J.~G. Andrews, ``Downlink cellular network analysis with
  multi-slope path loss models.,'' {\em IEEE Trans. Communications}, vol.~63,
  no.~5, pp.~1881--1894, 2015.

\bibitem{access2010further}
3GPP, ``Technical specification group radio access network; evolved universal
  terrestrial radio access (e-utra); further advancements for e-utra physical
  layer aspects (release 9). tr 36.814,'' tech. rep., 2010.

\bibitem{al2014optimal}
A.~Al-Hourani, S.~Kandeepan, and S.~Lardner, ``Optimal lap altitude for maximum
  coverage,'' {\em Wireless Communications Letters, IEEE}, vol.~3, no.~6,
  pp.~569--572, 2014.

\bibitem{stoyan}
S.~N. Chiu, D.~Stoyan, W.~S. Kendall, and J.~Mecke, {\em Stochastic geometry
  and its applications}.
\newblock John Wiley \& Sons, 2013.

\bibitem{sun2016propagation}
S.~Sun, T.~S. Rappaport, S.~Rangan, T.~A. Thomas, A.~Ghosh, I.~Z. Kovacs,
  I.~Rodriguez, O.~Koymen, A.~Partyka, and J.~Jarvelainen, ``Propagation path
  loss models for 5g urban micro-and macro-cellular scenarios,'' in {\em
  Vehicular Technology Conference (VTC Spring), 2016 IEEE 83rd}, pp.~1--6,
  IEEE, 2016.

\bibitem{active_u_yu2013downlink}
S.~M. Yu and S.-L. Kim, ``Downlink capacity and base station density in
  cellular networks,'' in {\em Modeling \& Optimization in Mobile, Ad Hoc \&
  Wireless Networks (WiOpt), 2013 11th International Symposium on},
  pp.~119--124, IEEE, 2013.

\bibitem{gil1951note}
J.~Gil-Pelaez, ``Note on the inversion theorem,'' {\em Biometrika}, vol.~38,
  no.~3-4, pp.~481--482, 1951.

\bibitem{dist_rec}
R.~M. Mnatsakanov and A.~S. Hakobyan, ``Recovery of distributions via
  moments,'' {\em Lecture Notes-Monograph Series}, pp.~252--265, 2009.

\end{thebibliography}
\end{document}